\documentclass[twocolumn,prb,floatfix,amsmath,amssymb,aps,superscriptaddress]{revtex4}
\usepackage{graphicx}

\begin{document}

\hsize\textwidth\columnwidth\hsize\csname@twocolumnfalse\endcsname

\title{Anharmonicity in one-dimensional electron-phonon system}

\author{Jize Zhao}
\affiliation{Institute for Solid State Physics, University of Tokyo,
Kashiwa, Chiba 277-8581, Japan}
\affiliation{Department of Physics, Renmin University of China, Beijing 100872, China}

\author{Kazuo Ueda}
\affiliation{Institute for Solid State Physics, University of Tokyo,
Kashiwa, Chiba 277-8581, Japan}

\date{\today}

\begin{abstract}
We investigate the effect of anharmonicity on the one-dimensional
half-filled Holstein model by using the determinant quantum Monte Carlo method.
By calculating the order parameters we find that with and without anharmonicity
there is always an transition from a disorder phase to a dimerized phase.
Moreover, in the dimerized phase a lattice dimerization and a charge density wave coexist.
The anharmonicity represented by the quartic term suppresses
the dimerization as well as the charge density wave, while a double-well
 potential favors the dimerization.
In addition, by calculating the correlation exponents we show that
the disorder phase is metallic with gapless charge excitations
and gapful spin excitations while in the dimerized phase both
excitations are gapful.
\end{abstract}

\maketitle

\section{Introduction}
The interaction between electrons and ion vibrations is responsible for 
many fundamental phenomena in solids. For example, as shown in BCS theory in
electron-phonon(EP) interaction induces a weak attraction between electrons near Fermi
surface with opposite momenta and opposite spins and eventually leads to superconductivity\cite{MAHAN1}. 
In quasi one-dimensional materials it is well known that EP 
coupling usually causes metal-insulator transition.
In the insulating phase a band gap opens at the Fermi energy and simultaneously lattice is distorted resulting 
in a larger unit cell. This transition is called Peierls transition\cite{PEIERLS1}. 
 
For simplicity most of theoretical works consider only harmonic ion vibration 
and in only a few works higher orders are taken into account.  
For example, the effect of anharmonicity on superconductivity was studied by 
Freericks et al.\cite{FREERICKS1} more than ten years ago in infinite-dimensional limit. However, it seems that the 
anharmonicity represented by the quartic term does not enhance the
superconductivity transition temperature as expected.
Another example is that polaronic properties with anharmonicity were investigated by  
Chatterjee and Takada\cite{CHATTERJEE1} some time ago.
 
Recently a renewed interest in anharmonic ion vibration was raised in 
$\beta$-pyrochlore oxides, especially in KOs$_2$O$_6$. 
Experiments show that it has unusual behaviors in both normal 
state and superconductivity state\cite{YONEZAWA1,YONEZAWA2, YONEZAWA3,
HIROI1,YAMAURA1}.  
In KOs$_2$O$_6$ an Os-O network forms a oversized cage and 
K ion sits in the cage. As the mass of K ion is small compared with Rb or Cs it oscillates 
relatively far from the equilibrium position. Density functional theory shows 
that this movement is highly anharmonic and the harmonic term of the potential 
can even be zero or negative\cite{KUNES1, KUNES2}.   
Some of the anomalous behaviors observed in KOs$_2$O$_6$ was explained by Dahm and one of
the present authors by treating the anharmonic ion vibration 
in a quasi-harmonic approximation with temperature-dependent frequency. \cite{DAHM1,DAHM2}. 
The spectral function and NMR relaxation rate have been discussed for a model with a general 
anharmonic potential\cite{TAKECHI1}. Otsuka et al\cite{OTSUKA1} 
has consider the effect of static anharmonic potential on quasi-one-dimensional extended Hubbard model and rich 
phase diagrams have been found at finite temperature. However, in quasi-one-dimensional materials,
phonon fluctuation is usually important\cite{LU1}.  

In this paper we will use the one-dimensional spinfull Holstein model as
a prototypical model to study the effects of anharmonicity of ion vibrations. 
Since here we are more interested in the anharmonicity the model we will study is given by 
\begin{eqnarray}
&&\mathcal{H}=\sum_{i}\frac{p_i^{2}}{2m}+V(r_{1}r_2\cdots r_{L})-t\sum_{i\sigma}(c^{\dagger}_{i\sigma}c_{i+1\sigma}+h.c.)\nonumber \\
&&~~~~~~~~~~-\lambda\sum_{i\sigma} r_{i}(n_{i\sigma}-\frac{1}{2})-\mu\sum_{i\sigma}n_{i\sigma} \label{HAM1}
\end{eqnarray} 
where $p_i$ and $r_i$ are the momentum and position operator of ions at site $i$, 
$V(r_1r_2\cdots r_L)=\sum_{i=1}^{L}(\frac{K}{2}r^2_i+\gamma{r^4_i})$ is the potential 
energy of ion vibration with $L$ being the lattice size. $\gamma$ is always nonnegative so that 
the ion potential is bounded. $t$ is the hopping term of electrons onto nearest-neighbor 
sites. $\sigma$ is the spin index and in our work 
we only consider spinful model and then $\sigma$ takes two values $\sigma=\uparrow,\downarrow$. 
$c^\dagger_{i\sigma}(c_{i\sigma})$ is the creation(annihilation)
operator for an electron at 
site $i$ with spin index $\sigma$, $n_{i\sigma}=c^\dagger_{i\sigma}c_{i\sigma}$ is the electron 
number operator at site $i$ for spin $\sigma$. The electrons couple locally with ions and the 
coupling constant is given by $\lambda$. $\mu$ is the chemical potential for controlling the electron number. 
If $\mu=0$ Hamiltonian $(\ref{HAM1})$ is invariant under the particle-hole transformation 
$p_i\rightarrow -p_i,r_i\rightarrow -r_i,c_i\rightarrow
(-1)^{i}c^\dagger_{i}$ 
and thus it is always half-filled at $\mu=0$. 

If $\gamma=0$ Hamiltonian (\ref{HAM1}) reduces to the {\it{standard Holstein model}}. 
In the past several decades it has been extensively studied by many authors. The mean-field theory and the  
expansion from the strong coupling limit predict that at half-filling for 
spinless case there is a phase transition from a disorder phase to a dimerized phase 
while for spinful case the ground state is dimerized for any 
finite EP coupling and any finite phonon frequency\cite{HIRSCH1}. 
However, the latter conclusion was challenged by Wu et al\cite{Wu1}. Based on functional integral analysis they argued 
that the fluctuation of phonons could also destroy the dimerization for spinful case when  
$\lambda$ is smaller than a critial $\lambda_c$. This conclusion was confirmed later by  
accurate density-matrix renormalization group calculations\cite{JECKELMANN1}
as well as quantum Monte carlo simulations\cite{HARDIKAR1}. 

However, it is not clear yet what is the effect of anharmonicity $\gamma\ne 0$ in one dimension. 
In particular, if the harmonic term $K$ is zero or even negative, the ion potential 
becomes double-well. Then a natural question is whether there are some fundamental changes compared
with the harmonic case? 
In the following we will clarify these issues by investegating the
Hamiltonian (\ref{HAM1}) by using the determinant 
quantum Monte Carlo simulations\cite{BLANKENBECLER1, SANTOS1}. In this model the electrons of two  
spin species couple equally to ions, so there is no sign problem.
In our simulations the largest lattice size is up to 32 and finite-size scaling is 
used to extrapolate our results to the thermodynamic limit. The inverse temperature $\beta$ is up to
24. We check our results with different $\beta$ to
confirm that the temperature in our simulation is low enough to extract ground state properties. 
The time slice $\Delta \tau=0.1$ is used in most of our simulations and we also check the results 
carefully by calculations with smaller $\Delta \tau$  and confirm that the results 
are not sensitive to $\Delta \tau$.  

In this paper we consider only the half-filled case by putting $\mu=0$. Moreover for simplicity we also
fix $m=0.5$ and $t=1$. In Section II,
firstly we will take $\omega=\sqrt{K/m}=1$, and show the results as a function of $\lambda$.
We find that for both $\gamma=0$ and finite $\gamma$ there is a phase transition 
from the disordered phase to the dimerized phase. The dimerization occurs  
accompanied by the charge density wave(CDW) transition simultaneously. 
Secondly we show the phase diagram in $(\lambda, \gamma)$ plane.
Finally we show the order parameters $m_p$ and $m_e$ as a function of 
$K$ with fixed $\lambda=1.4$ and $\gamma=0.1$. Similarly, a 
transition from the dimerized CDW phase to the disordered phase is found. 
In Section III, we will present results on static charge structure factor and spin structure factor.
These data confirm our previous conclusions. By calculating correlation exponents, we find that 
for any finite $\lambda$ there is a spin gap. However, a charge gap opens only after 
$\lambda>\lambda_c$, or $K<K_c$. In Section IV, we summarize our results.

\section{Dimerization and charge density wave order parameters}
Since Hamiltonian (\ref{HAM1}) with $\gamma=0$ has been extensively studied 
then we will focus on the case with a finite $\gamma$ in this work.  Data for $\gamma=0$ 
are calculated just for comparison with the DMRG results\cite{JECKELMANN1}. 
When $\gamma$ is larger than zero the anharmonic term increases the elastic energy of ion. 
Therefore a straightforward conjecture is that it will suppress the dimerization as well as the CDW order. 
In the first step we examine how the order parameters change as a function of $\lambda$ 
when the anharmonic term $\gamma$ is introduced. For our purpose we define the 
staggered correlation function of lattice displacements
\begin{eqnarray}
D_p(l)=\frac{1}{L}\sum_{i}(-1)^{l}\langle r_{i}r_{i+l}\rangle \label{DPL}
\end{eqnarray}
and the staggered correlation function of the electron densities
\begin{eqnarray}
D_e(l)=\frac{1}{L}\sum_{i}(-1)^{l}\langle n_{i}n_{i+l}-1\rangle \label{DEL}
\end{eqnarray}
The order parameters $m_p$ and $m_e$ are correspondingly defined as 
\begin{eqnarray}
m_{p}^{2}=\frac{1}{L}\sum_{l}D_p(l) \label{MP}
\end{eqnarray}
and 
\begin{eqnarray}
m_{e}^{2}=\frac{1}{L}\sum_{l}D_e(l) \label{ME}
\end{eqnarray} 
In the thermodynamic limit $m_{p}$ is expected to be finite in the dimerized 
phase while it is zero in the disorder phase. Similarly, $m_e$ is finite
in the 
CDW phase and zero in the disorder phase.
\begin{figure}
\includegraphics[width=6cm, height=7cm, clip]{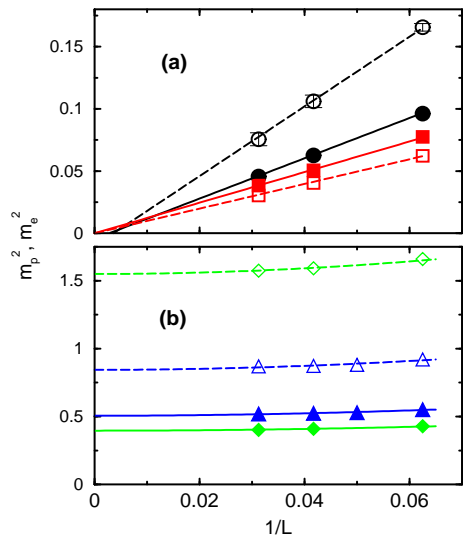}
\null \vskip-3mm
\caption{(Color online)Finite-size scaling of square of the order parameters $m_p^{2}$(open symbols) and $m_e^{2}$(filled symbols).
The parameters $(\lambda, \gamma)$ for black circles, red squares, green diamonds and blue triangles are $(0.7,0.0), (1.0, 0.1), (1.0, 0.0)$
and $(2.0,0.1)$, respectively.}
\label{FIG1}
\vskip-2mm
\end{figure}
To be specific we first fix $K=0.5$. In Fig. \ref{FIG1} we show both 
$m_p^2$ and $m_e^2$ as a function of inverse length of chain with various $(\lambda,\gamma)$. 
Now let us see the data without anharmonicity, i.e., $\gamma=0$.
As shown by circles in (a) both $m_p^2$ and $m_e^2$ clearly extrapolate 
to zero in the thermodynamic limit for $\lambda=0.7$ while for
$\lambda=1.0$(shown by diamonds in (b)) the order parameters  
are finite. This fact tells clearly that there is at leat one phase transition at $\lambda_c$ satisfying $0.7<\lambda_c<1.0$. 
Next we turn to see the data with $\gamma>0$. As a representative example, here we 
choose $\gamma=0.1$. As $\lambda=1.0$(shown by squares in (a)) both $m_p^2$ and $m_e^2$ vanish in the thermodynamic limit,
which is different from $\gamma=0$ case. However, as we increase EP
coupling $\lambda$ we reenter in the  
dimerized CDW phase, which is shown by triangles in Fig. \ref{FIG1}(b) with $\lambda=2$.
As we have seen, a larger $\lambda_c$ is required when the anharmonicity
$\gamma$ is present and this is consistent with our conjecture.  

To extract the critical point $\lambda_c$ accurately we calculate the order parameters 
by scaning $\lambda$ with fixed $\gamma$.
The order parameters as a function of $\lambda$ are shown for both 
\begin{figure}
\includegraphics[width= 7cm, clip]{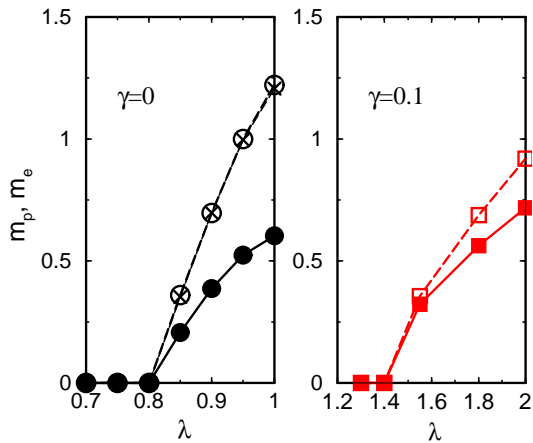}
\null \vskip-4mm
\caption{$m_p$(open symbols) and $m_e$(filled symbols) as a function of $\lambda$ for $\gamma=0$(left) and $\gamma=0.1$(right).
The crosses in the left panel are $m_e\times \lambda/K$. }
\label{FIG2}
\vskip-3mm
\end{figure}
$\gamma=0$ in the left panel of Fig. \ref{FIG2} and $\gamma=0.1$ in right panel of Fig. \ref{FIG2}, respectively. 
For $\gamma=0$ the data obtained by the QMC are in good agreement with
those obtained by the DMRG\cite{JECKELMANN1}.
The common feature for both $\gamma=0$ and $\gamma=0.1$ is that 
both $m_p$ and $m_e$ vanish at the same critical points, and thus there is one critical point. 
The critical values $\lambda_c$ are estimated to be 0.8 and 1.4, respectively.
For $\gamma=0$, $m_p$ and $m_e$ is found numerically\cite{JECKELMANN1}to satisfy
\begin{eqnarray}
m_p=\frac{\lambda}{K}m_e \label{MPME}
\end{eqnarray}
However, this relation does not hold for $\gamma=0.1$. In particular, $m_p/m_e$ is smaller than $\lambda/K$. 
It is plausible to say that the anharmonic 
potential $\gamma$ has "stronger" suppression on the ion dimerization than on the CDW formation.  
In Fig. \ref{FIG3} we show the   
schematic phase diagram in $(\lambda,\gamma)$ plane for $K=0.5$. We find that as $\gamma$ increases the critical value $\lambda$ also 
increases. This results also support our conclusion that the
anharmonicity suppresses the dimerization and CDW order.  
\begin{figure}
\includegraphics[width= 7cm, clip]{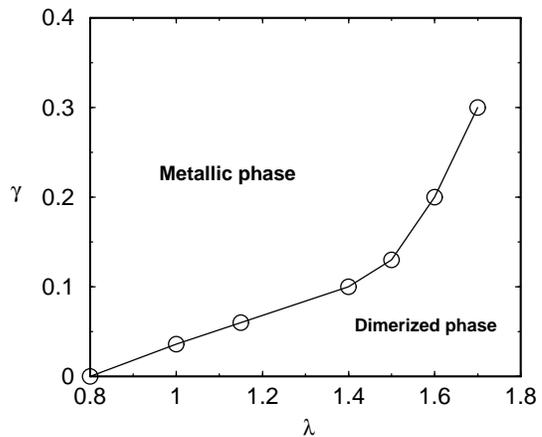}
\null \vskip-4mm
\caption{(Color online) Phase diagram in $(\lambda,\gamma)$ plane for $K=0.5$. The solid line is a guide of eyes.}
\label{FIG3}
\vskip-3mm
\end{figure}

\begin{figure}
\includegraphics[width= 7cm, clip]{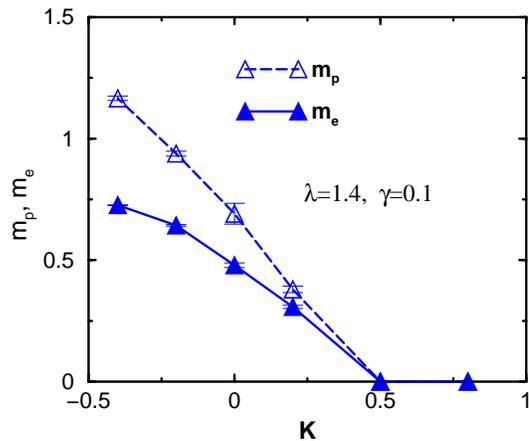}
\null \vskip-4mm
\caption{(Color online)Order parameters $m_p$ and $m_e$ are shown as a function of $K$. Here $\lambda$ is set to be 1.4 and $\gamma=0.1$.}
\label{FIG4}
\vskip-3mm
\end{figure}

As the coefficient of the harmonic term of the potential may be negative and then the total ion
potential has a form of a double-well,
we also calculate the order parameters as a function of $K$ with $\lambda=1.4$ and $\gamma=0.1$,  
which are shown in Fig. \ref{FIG4}.  
In this figure we see that when $K<K_c=0.5$ the ground state is in the dimerized CDW phase
while for $K>K_c$ it is in the disorder phase. No new 
phases are found in our simulations. Thus we may 
conclude that negative $K$ favors the dimerization with the CDW order
but does not introduce new phases in this model in one dimension.  

\section{Static structure factor and correlation exponents}
One-dimensional correlated electronic systems can usually be described by Tomonaga-Luttinger liquids(TLL)\cite{VOIT1}.
In the TLL theory charge degree of freedoms and spin degree of freedoms are
seperated. In particular, decay of the charge and spin correlation functions are determined 
by two exponents $K_\rho$ and $K_\sigma$, respectively.  
To explore properties of the disorder phase and the dimerized one further we calculate 
the charge structure factor and spin structure factor, which are defined by 
\begin{eqnarray}
\mathcal{C}(q)=\frac{1}{L}\sum_{ij}e^{iq(i-j)}(\langle n_i n_j\rangle-1).   \label{CK}
\end{eqnarray}
and
\begin{eqnarray}
\mathcal{S}(q)=\frac{1}{L}\sum_{ij}e^{iq(i-j)}S_i^z S_j^z.  \label{SK}
\end{eqnarray}
In particular, $K_\rho$ and $K_\sigma$ can be calculated\cite{CLAY1,HARDIKAR1} by 
\begin{eqnarray}
K_{\rho}=\pi\lim_{q\rightarrow 0}d\mathcal{C}(q)/dq. \label{KRHO}
\end{eqnarray}
and
\begin{eqnarray}
K_\sigma=4\pi \lim_{q\rightarrow 0}d\mathcal{S}(q)/dq, \label{KSIGMA}
\end{eqnarray}
\begin{figure}
\includegraphics[width= 7cm, clip]{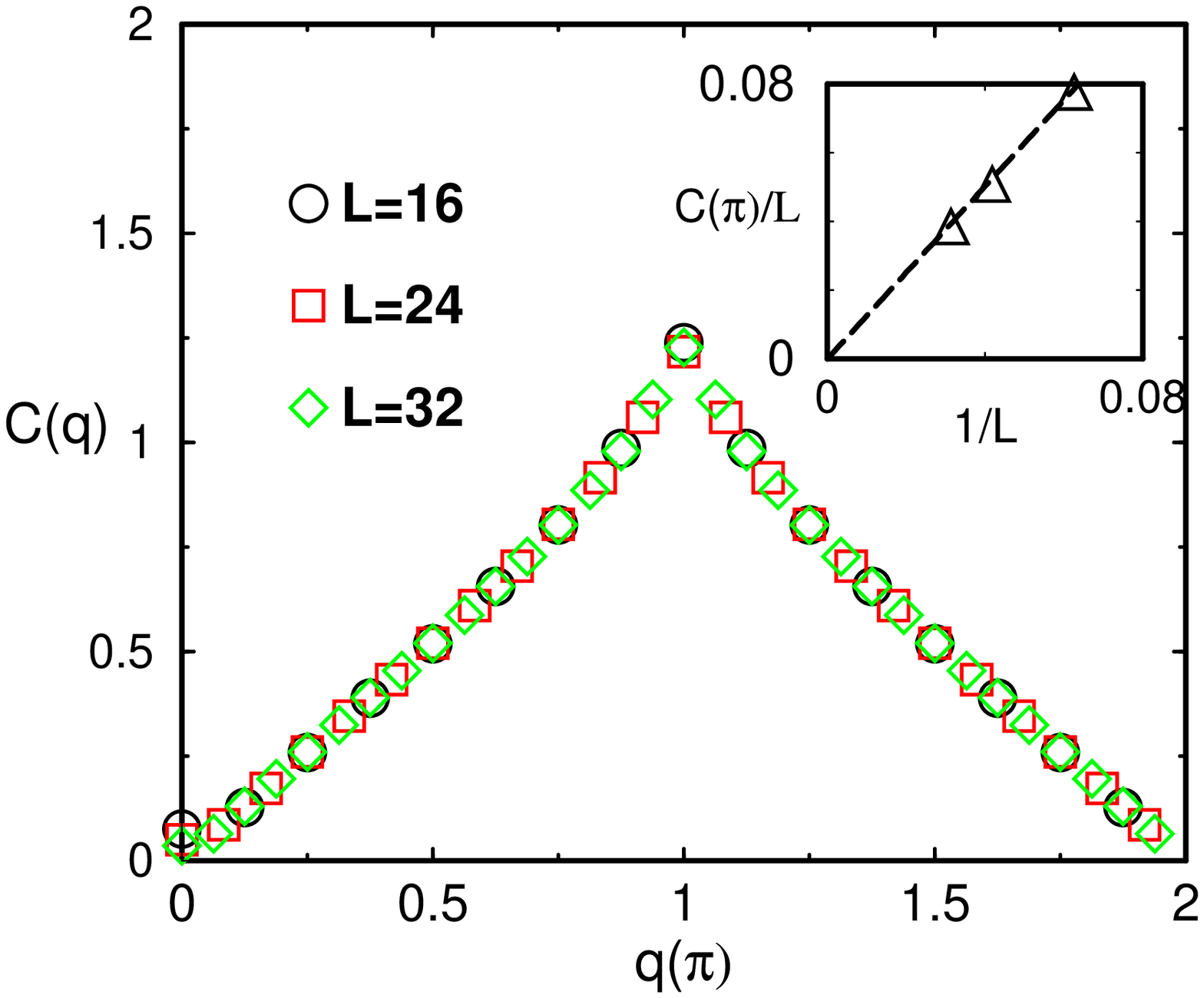}
\includegraphics[width= 7cm, clip]{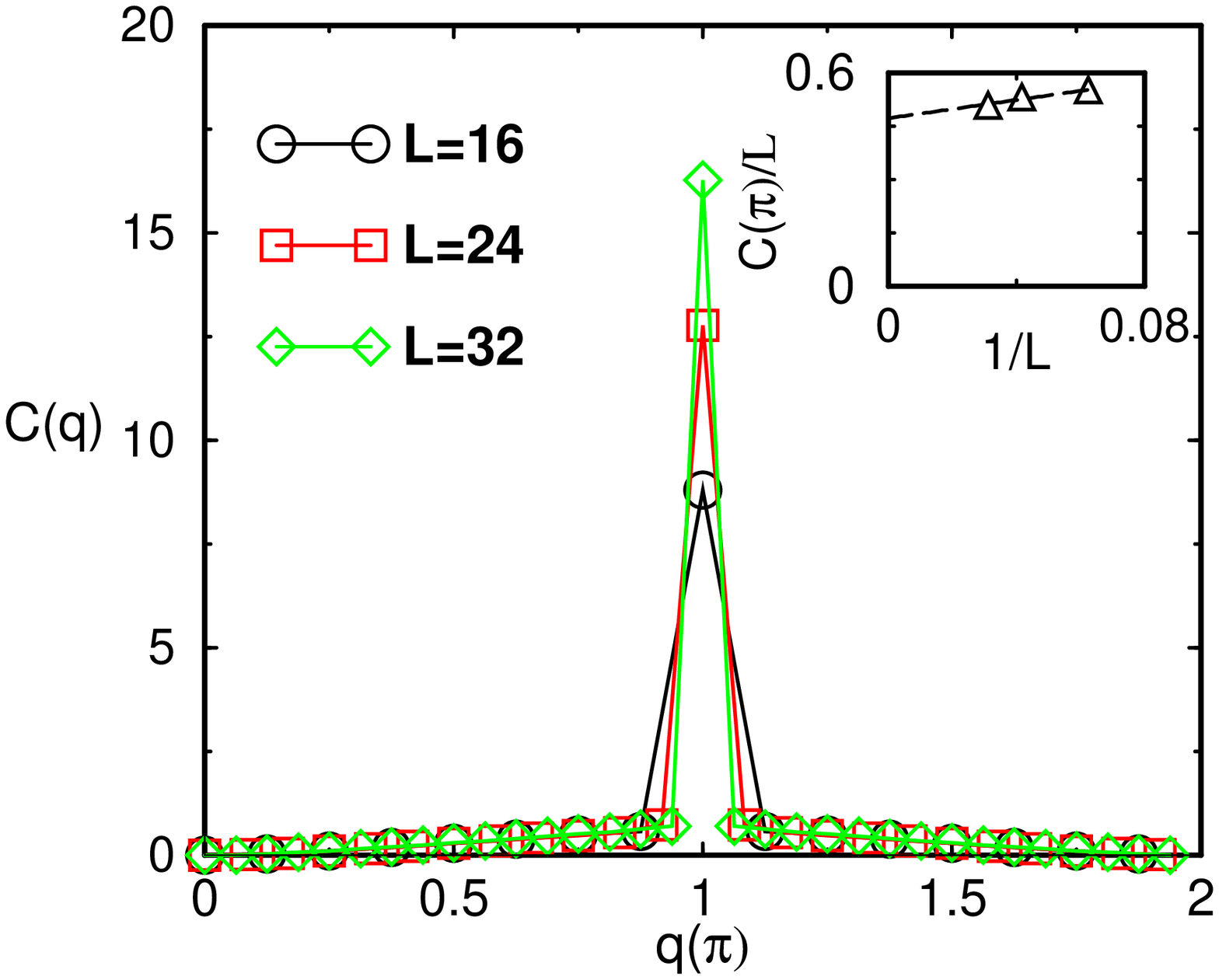}
\null \vskip-4mm
\caption{(Color online) Top: Static charge structure factor $\mathcal{C}(q)$ for $\lambda=1.0$. 
                             Inset: $\mathcal{C}(\pi)/L$ is shown as a function of $1/L$.
                        Bottom: Static charge structure factor for $\lambda=2.0$.
                             Inset: $\mathcal{C}(\pi)/L$ is shown as a function of $1/L$.}
\label{FIG5}
\vskip-3mm
\end{figure}
In Fig. \ref{FIG5} we show $\mathcal{C}(q)$ as a function of $q$ for $\lambda=1.0$ and $\lambda=2.0$ with 
$K=0.5$ and $\gamma=0.1$. 
In the top panel we find that $\mathcal{C}(q)$ for $L=16,24$ and $32$ fall almost into one curve and 
a peak is present at $q=\pi$. In the inset we show $\mathcal{C}(\pi)/L$ as a function of $1/L$. It vanishes 
in the thermodynamic limit.
This figure demonstrates that our sizes are large enough and finite size effect is negligible. 
One can conclude that there is only a dominant $2k_F$ short-range correlation and no long-range order.
In the bottom panel it is interesting to notice that $\mathcal{C}(q)$ exhibits a singularity 
at $q=\pi$ and $\mathcal{C}(\pi)/L$ is finite in the thermodynamic limit. 
This figure indicates that a long-range charge-charge correlation is present. 
These two figures are consistent with our conclusion that  $\lambda=1$
is in the disordered phase and $\lambda=2$ is in the dimerized CDW phase. 

Now we turn to study of the spin correlation functions.  
\begin{figure}
\includegraphics[width= 7cm, clip]{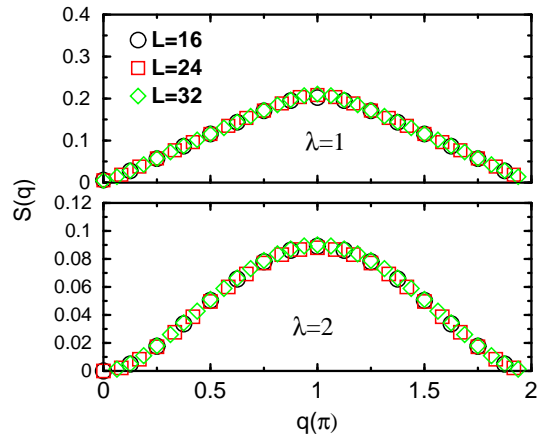}
\null \vskip-4mm
\caption{Static spin structure factors for $\lambda=1.0$ and $\lambda=2.0$ with the lattice size 
$L=16, 24$ and $32$. $K$ is chosen to be 0.5 and $\gamma$ is 0.1 and
 thus the parameter set used in the top panel 
is in the disorder phase and that of the bottom is in the dimerized phase. }
\label{FIG6}
\vskip-3mm
\end{figure}
In Fig. \ref{FIG6} we show the data of static spin structure factor for various lattice sizes
for $\lambda=1.0$ and $2.0$ at $K=0.5$ and $\gamma=0.1$. 
First, from the collapse of numerical data of various lattice sizes we may 
conclude that finite size effect is negalible in the data. 
Secondly we observe that a peak is present at $q=\pi$ for both figures. 
Moreover, the height of the peak is found to be finite in both $\lambda=1.0$ and $\lambda=2.0$. 
We then conclude that in this model there is a dominant short-range $2k_F$ spin correlation in both the disorder 
phase and the dimerized phase.

\begin{figure}
\includegraphics[width= 6cm, clip]{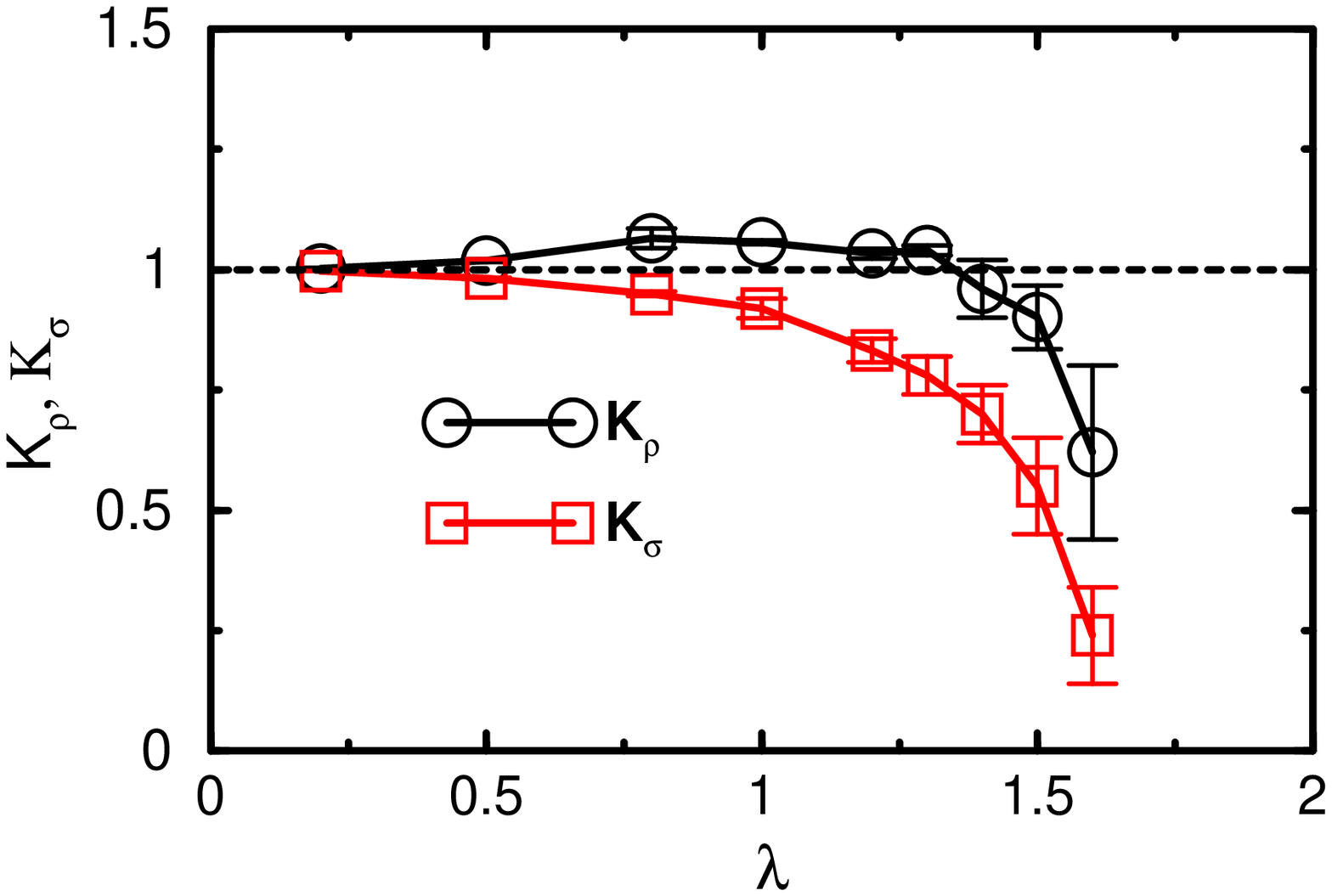}
\includegraphics[width= 6cm, clip]{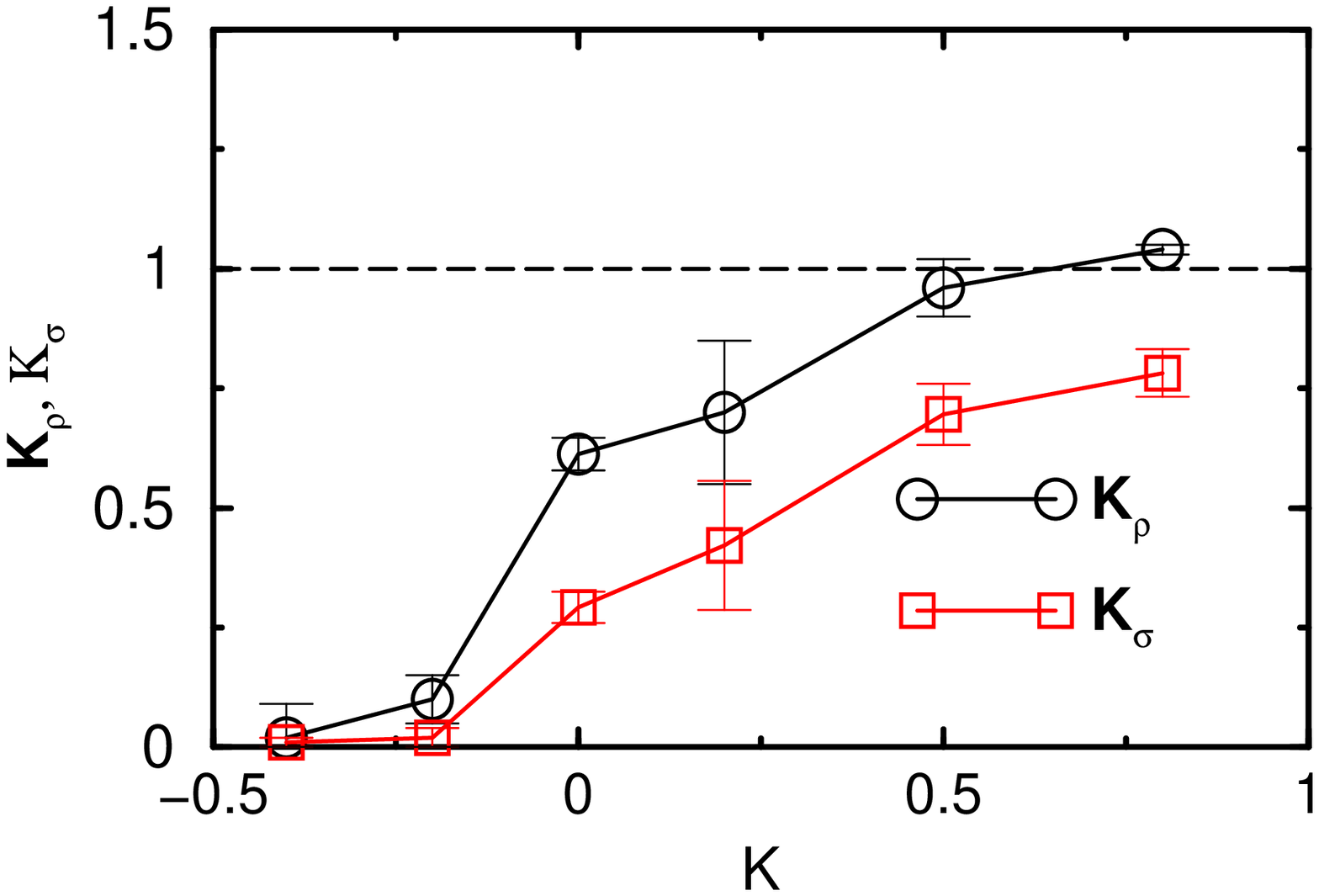}
\null \vskip-4mm
\caption{(Color online) Top: $K_\rho$ and $K_\sigma$ as a function of $\lambda$ for $K=0.5$ and $\gamma=0.1$.
                        Bottom: $K_\rho$ and $K_\sigma$ as a function of $K$ for $\lambda=1.4$ and $\gamma=0.1$.}
\label{FIG7}
\vskip-3mm
\end{figure}

Next we will analyse the charge and spin correlation exponents $K_\rho$  and $K_\sigma$. 
Since our model is invariant under spin rotation, we expect that if it
is in the TLL phase $K_\sigma$ is equal to 1.
However, if it is less than $1$ a spin gap is expected to open\cite{CLAY1,HARDIKAR1}.
This quantity is quite sensitive to the opening of a spin gap and therefore we use it 
as a criterion to determine whether there is a spin gap or not. 
In the top panel of Fig. \ref{FIG7} we show both $K_\rho$ and $K_\sigma$ as a function of 
$\lambda$ at $\gamma=0.1$. The behaviors are actually qualitatively similar 
to the results of the standard Holstein model, which are obtained by the
Monte carlo simulations \cite{HARDIKAR1}. 
Namely, in the weak coupling region, $K_\rho$ is larger than 1. Moreover, it shows nonmonotonous dependence 
on $\lambda$: it increases from 1 as $\lambda$ increases from zero, and after it reaches a maximum it starts to decrease. 
In the strong coupling region $K_\rho$ is always smaller than one.  
The crossover point with the line $K_\rho=1$ is estimated to be between $\lambda=1.3$ and $\lambda=1.4$.  
This value is in agreement with the critical point $\lambda_c=1.4$. We believe the small deviation is 
due to numerical errors. In the weak coupling region $K_\rho>1$ is intepreted as 
the attraction of effective electron-electron interaction mediated by
phonons while in the strong coupling 
region $K_\rho<1$ is intepreted as effective repulsive electron-electron
interaction in the dimerized phase.
However, $K_\sigma$ does not show such crossover and it always smaller than $1$ and 
monotonously decreases as a function of $\lambda$. We then conclude that a spin 
gap exists for any finite $\lambda$. 
In the bottom panel we show $K_\rho$ and $K_\sigma$ as a function of $K$ for $\gamma=0.1$. 
In this figure negative $K$ is also considered. Again $K_\rho$ crosses
from below 1 to above 1 as $K$ increases.  
The cross point seemingly is consistent with the critical point $K_c=0.5$ within our error bars.  
In the same way as before, $K_\sigma$ is small 
than $1$ in the whole range, indicating the presence of a spin gap. Compared with 
the results obtained by the DMRG\cite{JECKELMANN1} and the
QMC\cite{HARDIKAR1} we may conclude that the disorder phase is Luther-Emery liquid. 

\section{SUMMARIES}
In summary, in this paper we have investigated the effect of anharmonicity on the 
one-dimensional half-filled Holstein model. 
By studying order parameters, static charge and spin struture factors and correlation exponents we find 
a transition from Luther-Emergy liquid to dimerized CDW phase. This is essentially similar to standard Holstein model. 
The effect of anharmonicity is to suppress the dimerization as well as
the charge density wave. Recently, the effective 
interaction of electrons mediated by anharmonic phonons has been derived and the transition 
temperature of CDW of Hamiltonian (\ref{HAM1}) has been discussed\cite{DIKANDE1}. The transition temperature decreases 
monotonously as $\gamma$ increases. This behavior is also consistent
with the present conclusions. On the other hand, a double-well potential with a negative quartic term
favors the dimerization and the CDW order.

\acknowledgements
This work is supported by a Grant-in-Aid for Scientific Research on
Innovative Areas "Heavy Electrons" (No. 20102008) and also by Scientific
Research (C) (No. 20540347). J.Zhao is supported by part by the Japan Society for the Promotion of Science(P07036). 
Part of computation in this work has been done using the facilities of the Supercomputer Center, Institute for Solid State
Physics, University of Tokyo.

\end{document}